\documentclass[reprint,aps,prx,superscriptaddress,amsmath,amssymb,floatfix]{revtex4-2}
\usepackage{graphicx}
\usepackage{layout}
\usepackage{natbib}
\usepackage{dcolumn}
\usepackage{braket}
\usepackage{bm}
\usepackage{xcolor}
\usepackage{verbatim}
\usepackage{hyperref}
\usepackage{soul}
\usepackage[normalem]{ulem}

\setstcolor{red}

\begin{document}

\title{Fraxonium: Fractional fluxon states  for qudit encoding}

\author{Luca Chirolli}
\affiliation{Department of Physics and Astronomy, University of Florence, Via Sansone 1, IT-50019 Sesto Fiorentino, Italy}

\author{Valentina Brosco}

\affiliation{Istituto dei Sistemi Complessi, Consiglio Nazionale delle Ricerche, Via dei Taurini, 19 I-00185 Roma (IT)}
\affiliation{Physics Department,  University of Rome, ``La Sapienza'', P.le A. Moro, 2 I-00185 Roma (IT)}

\author{Uri Vool}
\affiliation{Max Planck Institute for Chemical Physics of Solids, 01187 Dresden (DE)}

\author{Gianluigi Catelani}
\affiliation{Quantum Research Center, Technology Innovation Institute, Abu Dhabi 9639, UAE}
\affiliation{Institute for Theoretical Nanoelectronics (PGI-2), Forschungszentrum J\"ulich, 52428 J\"ulich, Germany}

\author{Luigi Amico}
\affiliation{Quantum Research Center, Technology Innovation Institute, Abu Dhabi 9639, UAE}
\affiliation{Dipartimento di Fisica e Astronomia `Ettore Majorana', Via S. Sofia 64, 95123 Catania, Italy}
\affiliation{INFN-Sezione di Catania, Via S. Sofia 64, 95127 Catania, Italy}

\begin{abstract}
We propose a superconducting circuit hosting $d$ low-lying states, well separated from the rest of the spectrum, that naturally realizes a qudit system protected from leakage errors. The system represents a generalization of the  fluxonium and the low-energy states are constituted by fractional fluxon states, that we call {\it fraxons}, localized in the minima of a suitably designed Josephson potential. The latter is tailored through a Fourier engineering approach, that employs multi-harmonic Josephson building block elements composed by a Josephson junction and an inductance connected in series.  We present the spectrum of a $d=4$ and a $d=5$ qudit system and study in detail the
qutrit case. We analyze the dipole matrix elements for coupling to radiation and propose a non-Abelian, stimulated Raman adiabatic passage (STIRAP) protocol for single-qutrit gates, that is particularly suited for the present system.  The proposed platform opens novel perspectives in circuit engineering and quantum computing beyond the qubit paradigm.
\end{abstract}

\maketitle

\section{Introduction}

The pursuit of universal fault-tolerant quantum computing stands as one of the most ambitious goals  of quantum technology. While the field has witnessed remarkable progress in important features such as  gate fidelities and qubits coherence times, the transition to reliable, large-scale processors remains challenging. In this context,  quantum error correction  provides an important framework protecting logical information~\cite{shor1995,gottesman1997,gottesman2009}. Conventional quantum error correction protocols reach fault tolerance by encoding a single logical qubit across a substantial overhead of physical qubits. The embedding of qubits within topological constraints of complex networks has been extensively analyzed~\cite{kitaev2003,stern2013topological}, though this strategy would   typically be characterized by significant hardware overhead~\cite{fowler2012surface,google2023suppressing,bravyi2024high}.
An alternative paradigm to a more efficient computation involves expanding the {\it local} Hilbert space of the computational unit, so as to benefit from certain advantages in terms of algorithmic complexity.

Following the latter line of thought, here we will deal with  quantum systems  with  a {\it{finite number $d$  of levels}}, or qudits \cite{wang2020}. Indeed,  qudits offer a number of advantages in computational protocols, for example,  through  magic state distillation \cite{campbell2012, campbell2014}, compact encoding and increased space for information processing  \cite{kapit2016}, and reduction in circuit complexity \cite{gokhale2002,fedorov2012,ralph2007}. Quantum error correction schemes have been extended to qudits \cite{gottesman1999,muralidharan2017,grassl2018,majumdar2018,spencer2026}, and  
qudits have been proposed for optimal quantum circuits \cite{bullock2005}, improvement in Shor’s algorithm for factorization \cite{bocharov2017}, quantum neurons and Grover search \cite{gokhale2019}, quantum Fourier transform \cite{pavlidis2021}. In addition, they may offer an improvement in quantum cryptography \cite{bechmann2000,bruss2002,kaszlikowski2003}, through increased coding density and a higher security margin \cite{groblacher2006}, and quantum communication \cite{vaziri2002}, through a larger mutually unbiased basis \cite{hajji2022}.

\begin{figure}[t]
	\centering
    \includegraphics[width=1\linewidth]{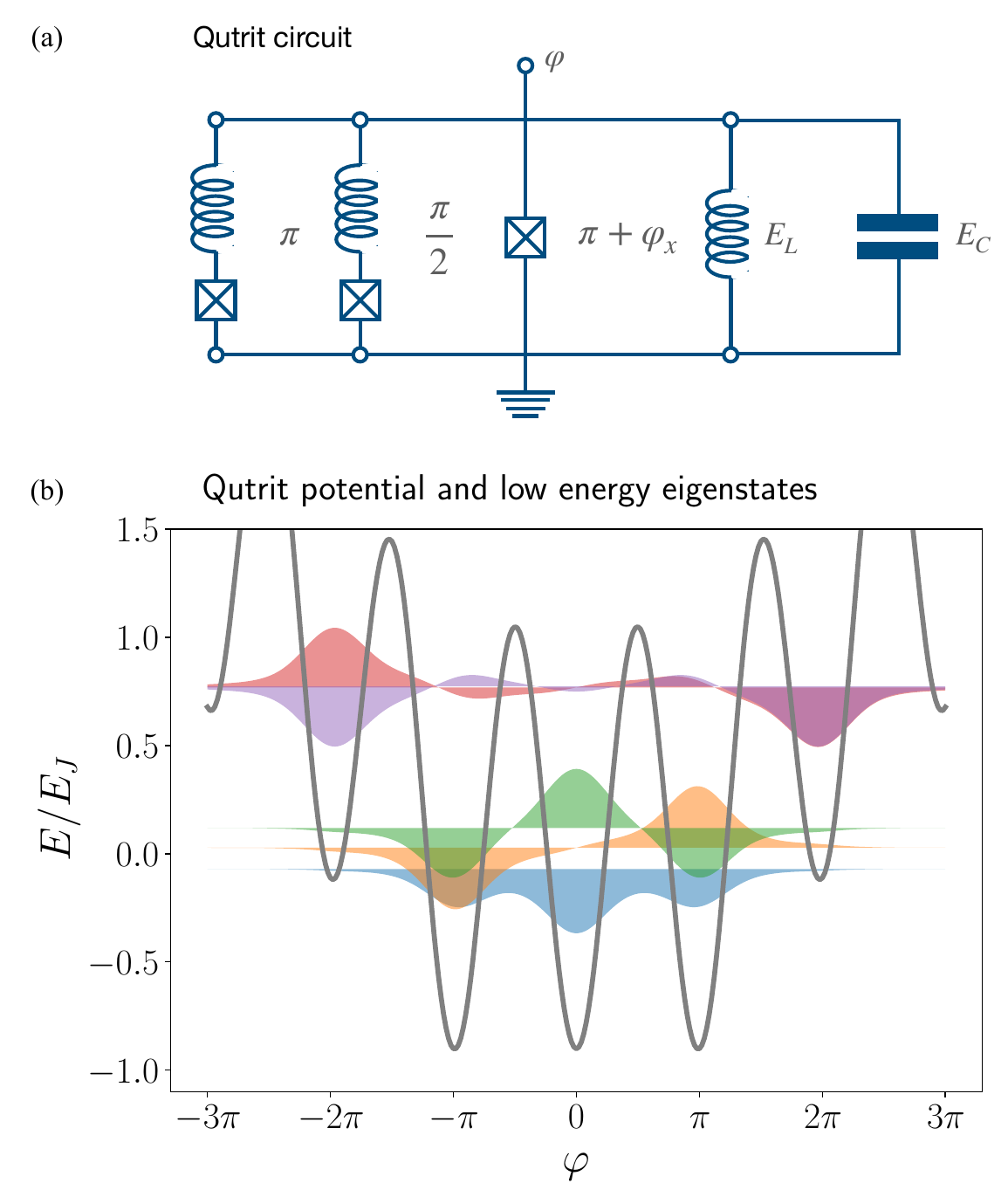}
	\caption{(a) Circuit realizing a qutrit composed by the following elements connected in parallel and with the fluxes threading the loops as shown: two nominally equal modular elements, each composed by an inductance in series with a Josephson junction and realizing a `kite' design with effective Josephson energy-phase relation $E_J\cos(2\varphi)$; a Josephson junction with energy-phase relation  $(\pi^2 E_L/4)\cos(\varphi)$; an inductance with energy $E_L$ and a capacitance of energy $E_C$.  The phase $\varphi$ is taken as a reference to drop across the single-harmonic Josephson junction and the fluxes threading the loops are as specified.   
    (b) Total generalized fluxonium potential realizing a qutrit and few eigenstates of the Hamiltonian Eq.~\eqref{Eq:Hqudit} with the Josephson potential Eq.~\eqref{Eq:qutrit-potential-general} for the case $E_C/E_J=0.15$ and $E_L/E_J=0.04$.}
	\label{fig:qutrit_main}
\end{figure}

In this work, we focus on superconducting circuits (for realizations of qudits in other platforms, see \cite{neves2005,lanyon2008,luo2019,imany2019,hu2020,chi2022,senko2015,ringbauer2022,hrmo2023,iqbal2025,lindon2023}). An important case in the context of expanding the local Hilbert space has been provided by the realization of error-corrected logical qudits \cite{brock2025} achieved through the Gottesman–Kitaev–Preskill (GKP) bosonic code \cite{gkp2001}.  This result extends previous achievements of qubit encoding via GKP states through superconducting cat states  \cite{sivak2023,campagne-ibarcq2020,grimm2020,grismo2021}. Noteworthy in this sense are the recent realization of a grid qubit \cite{nguyen2025} and the protocols employing superconducting cat-qubits \cite{Mirrahimi2014,lescanne2020exponential,Guillaud2019, chamberland2022building,Grimsmo2021}. Apart from these instances, physical qutrits in superconducting circuits are typically realized by employing the first three levels of the transmon \cite{shalibo2013,peterer2015,naik2017,cao2024},  and  single-   \cite{yurtalan2020,morovan2021,kononenko2021} and two-qutrit gates \cite{blok2021,goss2022,roy2023,goss2024,zhou2025,tripathi2025} have been demonstrated. However, the qutrit states are not well separated from the rest of the spectrum by a large energy gap, and the computation is therefore not protected against leakage errors outside the computational manifold.

In our approach, to protect the qudits computational states we draw inspiration from the fluxonium architecture, specifically leveraging its inherent inductive shunting and strong anharmonicity  \cite{manucharyan2009,nguyen2022,mcewen2021,reed2010,geerlings2013,suchara2015,magnard2018,bultink2020,varbanov2020,duan2021}; fast gates and long coherence times of fluxonium qubits have been demonstrated in Refs.~\cite{ficheux2021,xiong2022,somoroff2023}. 
Specifically, we propose a qudit design engineered to host $d$ degenerate fractional fluxon states, that we call {\it fraxons}; the latter  are quantum states of the phase localized in the minima of the effective potential and  corresponding to fractional flux quanta that can be trapped in the circuit.   Accordingly, we refer to our  resulting device as {\it fraxonium}. In previous work ~\cite{chirolli2021}, one of us showed that a \textit{flux qutrit} can be realized employing $\cos(2\varphi)$ elements; more generally, it was argued that Josephson engineering can enhance the flatness of the device spectrum and consequently its coherence properties. In the present work, we provide  circuits displaying $d$-degenerate minima of the total potential $U(\varphi)$; the latter is achieved by using higher harmonics of the Josephson terms to compensate the increase in energy due to the inductive term for the first $d$ minima. Fig.~\ref{fig:qutrit_main} explains the logic of our approach: the potential has been engineered so to have $d=3$ degenerate minima in a overall harmonic inductive envelope, thus generalizing the idea of the fluxonium and realizing a qutrit. We provide a general scheme of the circuit, valid for arbitrary $d$ values, by Fourier engineering the Josephson potential, which was proposed as  a systematic method of ``sculpting'' the potential energy of a superconducting circuit by controlling the individual harmonics of its energy-phase relation $E_{J,n}\cos(n\varphi)$ \cite{mert2023}. We shall see how indeed  high-harmonic elements can be realized by suitable parallel connections of modular elements composed of a series of a Josephson junction and an inductance, as proposed in what is known as  kite design \cite{brooks2013,smith2020,smith2022,smith2025,roverch2026}. 

The simplest of the high-harmonic terms is provided by a $\pi$-periodic element. The latter has been first proposed to arise in a circuit composed by four Josephson junctions in a SQUID configuration   \cite{blatter2001design,doucot2002pairing}. The main interest associated to $\pi$-periodic elements is the possibility to encode a parity-protected qubit, as proposed in Refs.~\cite{ioffe2002topologically,ioffe2002possible,doucot2003topological,doucot2005protected,kitaev2006protected,doucot2012physical} and experimentally realized in networks of Josephson junctions \cite{gladchenko2009superconducting,bell2014protected,bondar2025}, in the kite design comprising Josephson junctions and inductors~\cite{brooks2013,smith2020,smith2022,smith2025,roverch2026}, in hybrid semiconducting-superconducting devices~\cite{larsen2020parity-protected,messelot2024phase,arnault2025multiplet,leblanc2025gate,banszerus2024voltage-controlled,ciaccia2024charge-4e,banszerus2025hybrid}, and in twisted $d$-wave junctions \cite{brosco2024superconducting,coppo2024}. More generally, $\cos(2\varphi)$ elements open novel routes towards quantum computation with superconducting circuits, and they have been recently utilized in the implementation of a superconducting grid-state qubit~\cite{nguyen2025}. 

The work is structured as follows. In Sec.~\ref{Sec:General-qudit} we describe a general procedure for engineering arbitrary $d$ qudits; in Sec.~\ref{Sec:High-harmonic-Engineering} we provide a recipe for generating high-harmonic terms from lump elements such as an inductance and a Josephson junction; in Sec.~\ref{Sec:Fluxon-basis} we discuss the low-energy Hamiltonian describing fluxon dynamics and detail the ququart ($d=4$) and ququint ($d=5$) cases; in Sec.~\ref{Sec:qutrit}  we focus on the qutrit case, which requires only a $-\cos(2\varphi)$ element, and analyze its spectrum and effective low-energy model; in Sec.~\ref{Sec:qutrit-manipulation} we analyze the qutrit selection rules for coupling its levels to microwave radiation and propose a non-Abelian STIRAP protocol for single-qutrit gates.

\section{General qudit scheme}
\label{Sec:General-qudit}

We start by considering a scheme for realizing a generalized superconducting circuit featuring $d$ low-energy states, separated by a gap from higher-energy states. We consider a circuit featuring a capacitor $C$, an inductor $L$, and a Josephson element characterized by a $2\pi$-periodic potential energy $U_d(\varphi)$, all connected in parallel, such that the Hamiltonian has the general form
\begin{equation}\label{Eq:Hqudit}
H=-4E_C\partial_\varphi^2+\frac{E_L}{2}(\varphi-\varphi_x)^2+E_J U_d(\varphi),
\end{equation}
where $\varphi_x=2\pi \Phi_x/\Phi_0$ is a (normalized) flux threading the main inductance loop, with $\Phi_0=h/2e$ the superconducting flux quantum, $E_C=e^2/2C$ and $E_L=(\Phi_0/2\pi)^2/L$. For later use we also define the full potential $U(\varphi,\varphi_x)=E_L(\varphi-\varphi_x)^2/2+E_J U_d(\varphi)$. We assume $E_J$ to be the largest energy scale in the problem, $E_J\gg E_C,E_L$. The fluxonium case is obtained by employing a simple Josephson junction, such that the Josephson potential reads $U_2(\varphi)=-\cos(\varphi)$. The minima of the Josephson potential host fluxon states $|l\rangle$, with integer $l$, that are quantum states localized around the minima of $-\cos(\varphi)$ and describing current-circulating states with $l$ units of quantized flux $\Phi_0$. Threading half a flux quantum through the loop makes the states $|-l\rangle$ and $|l+1\rangle$ degenerate (or $|-l-1\rangle,|l\rangle$ for opposite sign of the inserted flux). In particular,  the lowest energy  states $|l=0\rangle, |l=1\rangle$ of the fluxonium are degenerate and the degeneracy is split by quantum phase slips, processes involving quantum tunneling through the barriers of the Josephson potential that are exponentially suppressed as $e^{-\sqrt{8E_J/E_C}}$ for tunneling between adjacent minima, yielding a splitting of order $\hbar\omega_p e^{-\sqrt{8E_J/E_C}}$, with $\omega_p=\sqrt{8E_CE_J}/\hbar$ the plasma frequency~\cite{manucharyan2009}; a more accurate estimate for the splitting can be obtained by taking into account renormalizations of $E_C$ and $E_J$ due to the inductive term~\cite{catelani2011}. Higher energy states correspond to either excitations within the wells of the Josephson potential, costing units of $\hbar\omega_p$, or to fluxon states with  higher $l$, with  energy cost $E_L(2\pi l-\pi)^2/2$. These two energy scales yield a gap in the energy spectrum that protects from leakage the two lowest states encoding a qubit. 

Our idea to implement qudits is to construct a generalization of the fluxonium potential in which the lowest-energy $d$ minima are degenerate. We shall see that the states localized around these minima are characterized by fractional values of the flux quantum and therefore they define  {\it fractional fluxon states} or {\it fraxons} \cite{chandran2003,tornes2005}.

\subsection{Construction of the qudit potential}

We start by describing a general procedure for the construction of the potential $U_d$ by superimposing arbitrary harmonic elements, such as $\cos(n\varphi)$ of which we assume to have full control. We first consider the case of odd $d$. The potential can be  written as
\begin{equation}
    E_JU_d(\varphi)=-E_J\cos(L_d\varphi)+\sum_{p_n<L_d} E_{J,n}\cos(p_n\varphi),
\end{equation}
for certain integers $L_d$ and $p_n$, with $p_n<L_d$, and energies $E_{J,n}$ to be determined. Assuming that $E_C,E_L,E_{J,n}\ll E_J$ and that $E_{J,n}\simeq E_L$, the minima of the potential will be located in close proximity of $2\pi l/L_d$ for integer $l$.  The mechanism we propose is to progressively add lower harmonics to compensate the increase of energy due to the inductive term $E_L(2\pi l/L_d)^2/2$.  This requirement amounts to imposing the degeneracy condition for the first $d$ minima of the full potential $U(\varphi)\equiv U(\varphi,\varphi_x=0)$ at $\varphi_x=0$, symmetrically located around $\varphi=0$. Due to the inversion symmetry $U(-\varphi)=U(\varphi)$ we are left with $(d-1)/2$ Fourier components, necessitating as many coefficients. By requiring all $\cos(2\pi p_nl/L_d)$ to have different values, that is guaranteed by any choice for which $0<p_n/L_d<1/2$, the most natural choice is $L_d=d-1$ and $p_n=n$, so that the desired potential reads
\begin{equation}\label{Eq:minimal_Ud}
    U^{\rm odd}_d=-\cos\left[(d-1)\varphi\right]+\frac{E_L}{E_J}\sum_{n=1}^{(d-1)/2}a_n\cos(n\varphi),
\end{equation}
where for practical purposes we rescaled the energies of the Josephson components $E_{J,n}$ by the inductive energy $E_L$, such that $E_{J,n}=E_L a_n$ and the small parameter $E_L/E_J\ll 1$ is manifest. The exact minima $\varphi_l$ of the potential $U(\varphi)$ will in general be slightly displaced from the minima of the leading term,  $\varphi_{l}=2\pi l/L_d+\delta\varphi_l$. For $E_L\ll E_J$ we have $\delta\varphi_l\sim E_L/E_J\ll 1$, and we can neglect the corrections $\delta\varphi_l$. Likewise, the extremal condition will be satisfied up to corrections of order $E_L/E_J$. 

For the case of even $d$, the procedure is analogous, with the main difference that the leading term has the form $\cos(L_d\varphi)$; then the simplest choice is $L_d=d$, such that
\begin{equation}\label{Eq:minimal_Ud_even}
    U^{\rm even}_d=\cos(d\varphi)+\frac{E_L}{E_J}\sum_{n=1}^{d/2-1}a_n\cos(n\varphi),
\end{equation}
with the minima located in close proximity of $\varphi_l=2\pi (l+1/2)/d$. By imposing degeneracy of the  first $d$ minima we obtain the values of the coefficients $a_n$ (see Appendix~\ref{App:potential_engineering}).

We now consider a few examples. For the qutrit case $d=3$, we only need two harmonic components. Equating the potential at $\varphi=0$ with the potential at $\varphi=\pi+\delta \varphi$, with $\delta \varphi\propto {\cal O}(E_L/E_J)$, and expanding in powers of $E_L/E_J$ we obtain
\begin{equation}\label{Eq:qutrit-potential-general}
    U_3(\varphi)=-\cos(2\varphi)+\frac{E_L}{E_J}\left(1-\frac{E_L}{4E_J}\right)\frac{\pi^2}{4}\cos(\varphi).
\end{equation}
and $\delta\varphi=-\pi E_L/4 E_J+{\cal O}((E_L/E_J)^2)$; it can also be shown that the extremal condition is satisfied at order $E_L/E_J$. The effective circuit and the full potential $U_3 (\varphi)$ for the qutrit are shown in Fig.~\ref{fig:qutrit_main}(a,b), respectively, where the first three minima appear to be degenerate to a very good approximation. Proceeding in an analogous way, for the cases $d=4$ and $d=5$ the potential reads
\begin{eqnarray}
    U_4(\varphi)&=&\cos(4\varphi)+\frac{\pi^2}{4\sqrt{2}}\frac{E_L}{E_J}\cos(\varphi),\label{Eq:U4}\\
    U_5(\varphi)&=&-\cos(4\varphi)+\frac{\pi^2E_L}{4E_J}\left[-\frac{1}{4}\cos(2\varphi)+\cos(\varphi)\right]\!. \quad \label{Eq:U5}
\end{eqnarray}
The potentials for the cases $d=3,\,4,\,5$ are shown in Fig.~\ref{fig:potential-qudit}(a).

\section{High-harmonic engineering}
\label{Sec:High-harmonic-Engineering}

\begin{figure}[t]
	\centering
    \includegraphics[width=1.\linewidth]{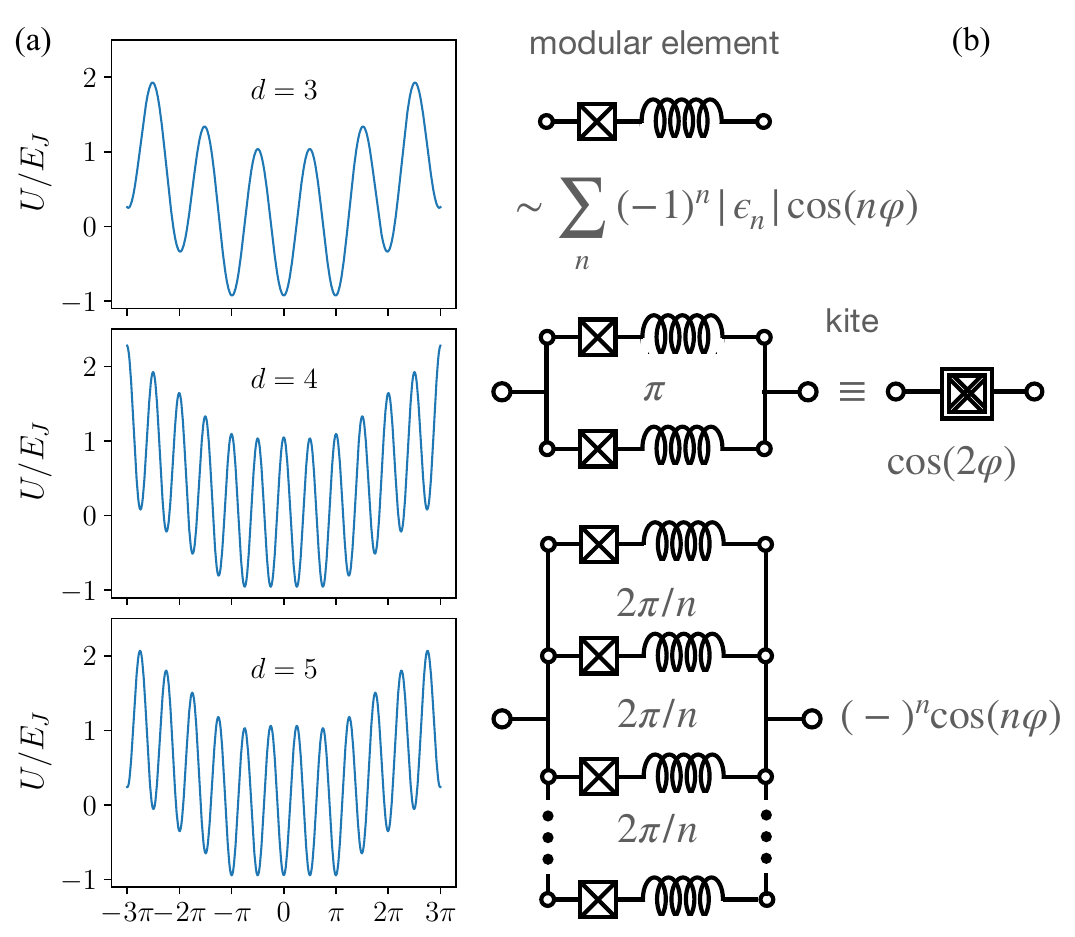}
	\caption{(a) Three instances of  potential $U(\varphi)$ featuring $d$ degenerate minima, for the cases $d=3,4,5$ and given by Eqs.~\eqref{Eq:qutrit-potential-general}, \eqref{Eq:U4}, \eqref{Eq:U5}, respectively.  (b) Scheme for obtaining high-harmonic elements with leading term $(-)^n\cos(n\varphi)$ obtained by connecting in parallel the modular element composed by a Josephson junction and an inductance, shown on the top of (b). The cases realizing a kite design, describing a $\cos(2\varphi)$ energy-phase relation, and the general $(-)^n\cos(n\varphi)$ are shown.}
	\label{fig:potential-qudit}
\end{figure}

Crucial to the construction of the qudit Hamiltonian is the availability of high-harmonic elements described by an effective $\cos(n\varphi)$ energy-phase relation. To this end, we rely on the idea at the basis of the so called `kite' circuit~\cite{brooks2013,smith2020,smith2022,smith2025}. Its building block [see Fig.~\ref{fig:potential-qudit}(b) top] is composed by a Josephson junction and an inductance connected in series, and it is described an energy-phase relation by $E_J(\varphi,\delta\varphi)=-E_J\cos[(\varphi+\delta\varphi)/2]+E_L(\varphi-\delta\varphi)^2/8$, with $\varphi$ the total phase drop across the element and $\delta\varphi$ the relative phase drop. Minimization with respect to $\delta\varphi$ yields an effective energy-phase relation $E_J(\varphi)$ that contains all harmonics. This can be explicitly seen for the case $E_J/E_L\ll 1$ by inserting the extremal solution $\delta\varphi^*=\varphi-(2E_J/E_L)\sin[(\varphi+\delta\varphi^*)/2]$ in $E_J(\varphi,\delta\varphi^*)$ and expanding it in powers of $E_J/E_L$. A similar approach, based on two junctions in series, was proposed in Ref.~\cite{mert2023} and experimentally realized in Ref.~\cite{shagalov2025}, but the presence of a floating island between the junctions makes the circuit sensitive to charge noise, a problem resolved by using an inductor instead. The kite design~\cite{brooks2013,smith2020,smith2022,smith2025,roverch2026} employs two junction-plus-inductor elements connected in parallel and threaded by a half flux quantum: by connecting in parallel two equal modular elements described by a generic energy-phase relation $E_J(\varphi)$, characterized by a high harmonic content, and inserting a half-flux quantum $\Phi_0/2=h/4e$ in the loop produces a $\pi$-periodic energy-phase relation, $E_{J\pi}(\varphi)=E_J(\varphi)+E_J(\varphi+\pi)$, since the $2\pi$-periodic contributions cancel out, as in the middle of Fig.~\ref{fig:potential-qudit}(b). A parity-protected qubit that employs the same rational with two gatemons has been realized~\cite{larsen2020parity-protected}. It is straightforward to generalize this approach: by 
connecting three modular elements in parallel and inserting a reduced flux $2\pi/3$ in each loop, $E_{J,2\pi/3}(\varphi)=E_J(\varphi)+E_J(\varphi+2\pi/3)+E_J(\varphi+4\pi/3)$, we obtain an object that is $2\pi/3$-periodic and whose dominant component is $-\cos(3\varphi)$. It is clear that this scheme can be iterated to generate all the desired harmonics [Fig.~\ref{fig:potential-qudit}(b) bottom].

\begin{figure*}[t]
	\centering
    \includegraphics[width=1\linewidth]{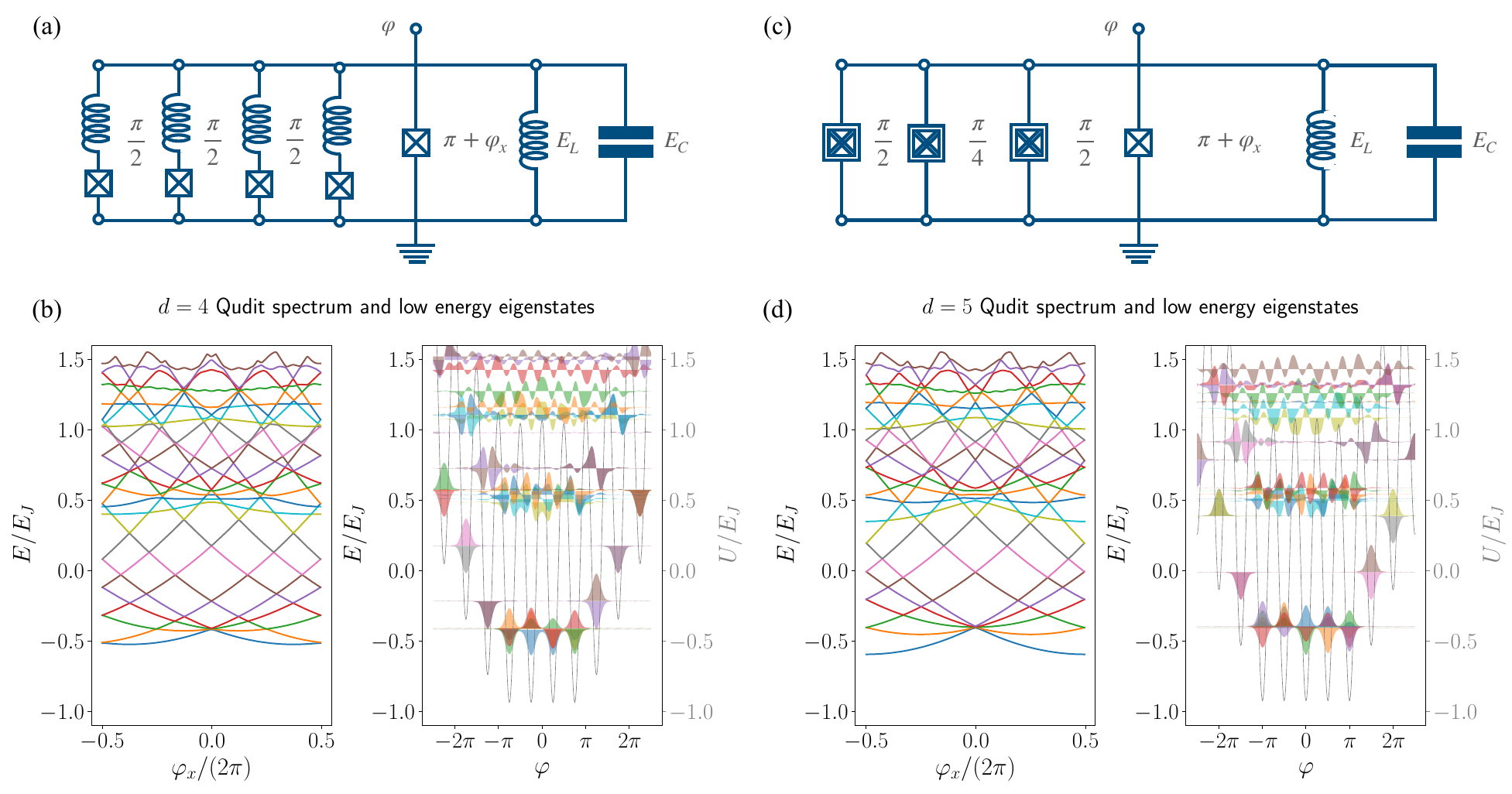}
	\caption{(a) Schematics of the circuit realizing the Hamiltonian by Eq.~\eqref{Eq:Hd4} and describing a $d=4$ qudit (ququart). It is  composed by a capacitance $C$, an inductance $L$, a Josephson junction with energy $E_L\pi^2/(4\sqrt{2})$ and four, nominally equal, modular elements--each composed by a Josephson junction and an inductance connected in series. All the branches are connected in parallel, with a flux $\pi/2$ between the four equal modular elements. The latter effectively yield 
    a $\pi/2$-periodic Josephson junction with energy $E_J$.  (b) Spectrum (left panel) and potential and few eigenfunctions (right panel) of the Hamiltonian Eq.~\eqref{Eq:Hd4} as a function of $\varphi_x$. (c) 
    Schematics of the circuit realizing the Hamiltonian by Eq.~\eqref{Eq:Hd5} and describing a $d=5$ qudit (ququint). The system features a Josephson junction with energy $E_L\pi^2/8$, a $\pi$-periodic Josephson junction (symbolized by a crossed-hatched square) with energy $E_L\pi^2/32$, and a $\pi/2$-periodic Josephson junction with energy $E_J$, all connected in parallel. The $\pi/2$-periodic element can be either obtain as in (a) or via two $\pi$-periodic elements connected in parallel and with a flux $\pi/2$ between the two.  
    (d) Spectrum (left panel) and potential and few eigenfunctions (right panel) of the Hamiltonian Eq.~\eqref{Eq:Hd5} as a function of $\varphi_x$. In (b) and (d) the parameters chosen are $E_C/E_J=0.01$ and $E_L/E_J=0.04$.}
	\label{fig:d4-d5}
\end{figure*}

\label{App:energy-phase-expansion]}
To be specific, the general expansion in Fourier harmonics of the Josephson potential describing the modular element [Fig.~\ref{fig:potential-qudit}(b) top] reads  $E_J(\varphi)=\sum_{n=1}^\infty (-1)^n|\epsilon_n|\cos(n\varphi)$ [see Appendix~\ref{App:energy-phase-expansion}]. It features alternating  signs between the even and odd harmonics~\cite{willsch2024,chirolli2025}, so that all even harmonics have a positive sign. The addition in parallel of $n$ terms with a flux $2\pi/n$ between adjacent elements produces the energy-phase relation  
\begin{equation}\label{Eq:parallel-EJ}
     \sum_{j=0}^{n-1}E_J(\varphi+2\pi j/n)\sim n(-1)^n |\epsilon_n| \cos(n\varphi),
\end{equation}
as schematized in Fig.~\ref{fig:potential-qudit}(b) bottom. Furthermore, the amplitude of the harmonics is suppressed by $\tau=E_J/E_L$, $\epsilon_n\propto \tau^n$, that behaves analogously to the transparency of the conduction channels through a Josephson junction \cite{willsch2024,chirolli2025}. It follows that the higher the harmonic order, the smaller the amplitude of the effective term; therefore, minimizing the order of the harmonics is the simplest and best choice.

In general we need to have the freedom to choose the sign of each harmonic. To this end, we notice that a way to obtain an overall minus sign in the effective element has been recently proposed \cite{hays2025}. It consists in placing two of the desired elements in series, so that the relative phase drops adjusts as to have zero total phase drop across the element. Indeed, the series of two equal $2\pi/n$-periodic elements reads $\cos(n(\varphi+\delta\varphi)/2)+\cos(n(\varphi-\delta\varphi)/2)=2\cos(n\varphi/2)\cos(n\delta\varphi/2)$ and minimization with respect to the relative phase $\delta\varphi$ yields $-2|\cos(n\varphi/2)|$, which is still $2\pi/n$-periodic and has the desired negative sign.
Finally, we point out that the scheme for generating higher harmonics is not unique and other choices of the fluxes piercing the loops could be employed.

\section{Fractional Fluxon basis}
\label{Sec:Fluxon-basis}

A low-energy Hamiltonian can be derived that describes the lowest $d$ energy states of the system and corresponds to the dynamics between the fraxons induced by the charging term. Because of the latter term, adjacent fraxons $|l\rangle$ can be coupled by  quantum phase slip processes (despite their fractional character, we still label fraxons through an integer index $l$). The fraxon states can be approximately described  with Gaussians of the form $\langle\varphi|l\rangle=e^{-(\varphi-\varphi_l)^2/2\ell_d^2}(\ell_d\sqrt{\pi})^{-1/2}$, with spread $\ell_d=(8E_C/L_d^2E_J)^{1/4}$, each well-localized around the minima $\varphi_l=2\pi l/L_d$ of the $2\pi/L_d$-periodic part of the potential. These states have an exponentially small but finite overlap, that can be neglected to a good approximation, $\langle l'|l\rangle\simeq \delta_{l,l'}$, and have energy
\begin{equation}\label{Eq:varepsilon_l}
    \varepsilon_l=\frac{E_L}{2}(\varphi_l-\varphi_x)^2-\frac{E_L}{2}\varphi_l^2+E_0,
\end{equation}
with $E_0$ weakly dependent on $l$ and $\varphi_x$ [the details can be found in Appendix~\ref{App:tight_binding}]. We see that the $\varepsilon_l$'s correspond to parabolas as a function of $\varphi_x$, centered around $\varphi_l$, and shifted in energy with respect to the value at $l=0$ by $-E_L\varphi_l^2/2$. For $E_L\ll\omega_p$, being $\omega_p=L_d\sqrt{8E_CE_J}$ the plasma frequency in each well of $\pm  E_J\cos(L_d\varphi)$  [with the $+(-)$ sign for for the even(odd)-$d$ case], we can neglect higher energy excitations and retain only  matrix elements $t_{l,l'}$ between fraxon states that account for quantum phase slips induced by the charging term. The effective Hamiltonian reads
\begin{equation}\label{Eq:Hfluxon}
    H_{\rm qudit}=\sum'_l\varepsilon_l|l\rangle\langle l|-\sum_{l,l'}t_{l,l'}|l\rangle\langle l'|+{\rm H.c.},
\end{equation}
where the primed sum is restricted to the $d$ quasi-degenerate fraxon states. Furthermore, due to the exponential localization of the fraxon states for $E_J\gg E_C$, we retain only the coupling between nearest-neighbor minima through the tunneling matrix element $t$. For the evaluation of $t$ we can employ standard results obtained through a WKB approximation \cite{catelani2011}, and the details can be found in Appendix~\ref{App:tight-binding}. When including only the $d$ degenerate fraxon states the Hamiltonian Eq.~\eqref{Eq:Hfluxon} well describes the low-energy sector of the system. In the following, we provide examples of the qudit spectrum for the cases of a ququart ($d=4$ qudit) and a ququint ($d=5$ qudit).

\subsection{Ququart and Ququint}

The full Hamiltonian of a ququart, that is a qudit with $d=4$, is given by
\begin{eqnarray}\label{Eq:Hd4}
    H_{d=4}&=&-4E_C\partial_\varphi^2+\frac{E_L}{2}(\varphi-\varphi_x)^2+E_J\cos(4\varphi)\nonumber\\
    &+&\frac{\pi^2}{4\sqrt{2}}E_L\cos(\varphi).
\end{eqnarray}
The circuit realizing the above Hamiltonian employs a $\cos(4\varphi)$ element which, as explained in Sec.~\ref{Sec:High-harmonic-Engineering}, can be realized by four nominally equal modular elements connected in parallel, with a $\pi/2$ flux inserted between adjacent elements, as shown in Fig.~\ref{fig:d4-d5}(a). Furthermore, the positive sign of the $\pi/2$-periodic element naturally arises from the scheme of Sec.~\ref{Sec:High-harmonic-Engineering}. The spectrum can be obtained numerically by expanding the Hamiltonian in the basis of the harmonic oscillator given by the capacitive and purely inductive term, as explained in Appendix~\ref{App:qutrit-exact-spectrum}, and it is shown as a function of the flux $\varphi_x$ in Fig.~\ref{fig:d4-d5}(b). The first four low-energy states correspond to the  fractional fluxon states localized around $\varphi_l=2\pi(l+1/2)/4$, for $l=-2,-1,0,1$, and their energies grow quadratically as a function of $\varphi_x$, in agreement with the expression for $\varepsilon_l$ provided in Eq.~\eqref{Eq:varepsilon_l}.  Around $\varphi_x=0$, they become approximately degenerate and weak quantum phase slip processes hybridize them.

Concerning higher-energy levels, two types of excited states appear in the spectrum at $\varphi_x=0$: {\it i)} pairs of fluxon states $(-l-1,l)$, which are shifted up in energy by approximately $E_L(\varphi_l)^2/2$ and hybridize very weakly  across the barriers of the potential for $\varphi_x=0$, in particular the pairs $l=(-3,2)$ and $l=(-4,3)$ appear as excited states above the ququart manifold and display a very weak avoided crossing [hardly appreciable in Fig.~\ref{fig:d4-d5}(b)]; {\it ii)}  plasmon states, that corresponds to excited states in the four wells of $\cos(4\varphi)$, higher in energy approximately by units of  $\omega_p=4\sqrt{8E_CE_J}$.

We next consider the case of a ququint, that is a qudit with $d=5$, and is described by the Hamiltonian   
\begin{eqnarray}\label{Eq:Hd5}
    H_{d=5}&=&-4E_C\partial_\varphi^2+\frac{E_L}{2}(\varphi-\varphi_x)^2-E_J\cos(4\varphi)\nonumber\\
    &+&\frac{\pi^2E_L}{8}(\cos(\varphi)-\frac{1}{4}\cos(2\varphi)).
\end{eqnarray} 
The circuit realizing this Hamiltonian features two even harmonic terms with negative sign. For nominally equal elements, it can be realized by employing three kite elements connected in parallel, as in Fig.~\ref{fig:d4-d5}(c), with the fluxes inserted as depicted. The first five low-energy states correspond to fractional fluxon states localized around $\varphi_l=\pi l/2$ with $l=0,\pm1, \pm2$, and with energy $\varepsilon_l$ given by Eq.~\eqref{Eq:varepsilon_l}. These states become approximately degenerate around $\varphi_x=0$, where quantum phase slip processes weakly hybridize them. As in the case of $d=4$, higher energy states are comprised by fluxon states with $|l|>2$, costing additional energy on order of $E_L(\varphi_l)^2/2$, and plasmon states excited in the wells of the $-\cos(4\varphi)$ potential, costing units of $\omega_p=4\sqrt{8E_CE_J}$. 

\begin{figure}[t]
	\centering
    \includegraphics[width=1\linewidth]{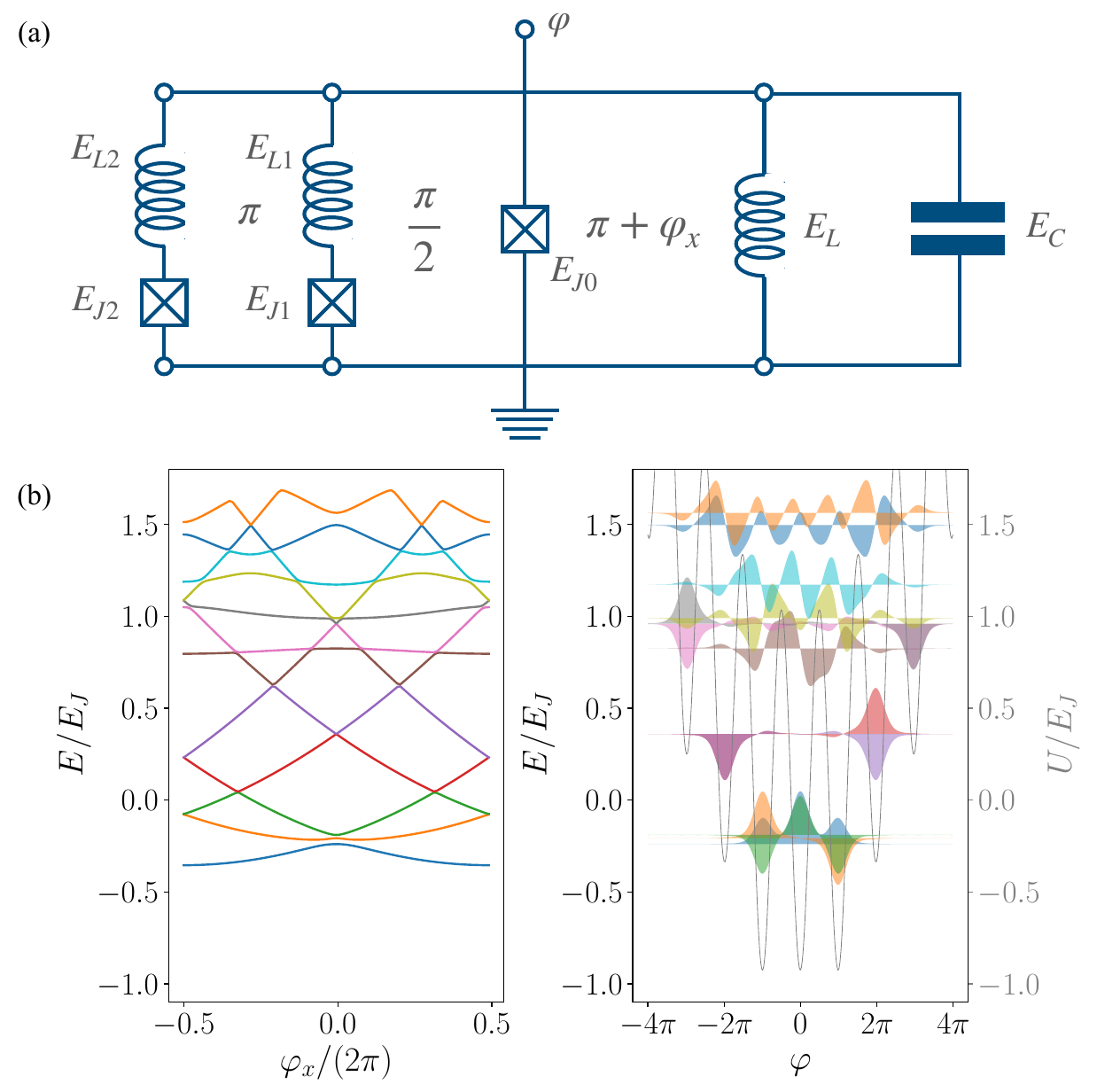}
	\caption{(a) Circuit realizing a qutrit. (b) Left panel: Spectrum of the Hamiltonian Eq.~\eqref{Eq:Hqutrit} for the choice $E_C/E_J=0.08$, $E_L/E_J=0.03$ as a function of $\varphi_x$. Right panel: Josephson potential and seventeen  eigenfunctions as a function of $\varphi$ (the color code matches the levels in the left panel).}
	\label{fig_qutrit_circuit}
\end{figure}

\section{Qutrit}
\label{Sec:qutrit}

We now focus on the case of a qutrit, that is realized by the circuit depicted in Fig.~\ref{fig:qutrit_main}(a), broken down into conventional elements in Fig.~\ref{fig_qutrit_circuit}(a). For the general case of different modular elements, as shown in Fig.~\ref{fig_qutrit_circuit}(a), the kite element is described by a potential of the form [see Appendix \ref{App:energy-phase-expansion}]
\begin{equation}\label{Eq:kite}
    U_{\rm kite}=-\tilde{E}_{J}\cos(\varphi)+\tilde{E}_K\cos(2\varphi),
\end{equation}
where $\tilde{E}_J$ and $\tilde{E}_K$ are functions of the input parameters $E_{J1}, E_{J2}, E_{L1}, E_{L2}$ \cite{smith2020}. Biasing the kite with flux $\pi/2$ to obtain a negative $\cos(2\varphi)$ component,
the circuit depicted in  Fig.~\ref{fig_qutrit_circuit} gives rise to the Hamiltonian
\begin{eqnarray}
\label{Eq:Hcircuit}
H&=&-4E_C\partial_\varphi^2+\frac{E_L}{2}(\varphi-\varphi_x)^2+E_{J0}\cos(\varphi)\nonumber\\
&-&\tilde{E}_J\sin(\varphi)-\tilde{E}_K\cos(2\varphi).
\end{eqnarray}
We clearly see that asymmetries in the kite circuit introduce a perturbation $\tilde{E}_J\sin(\varphi)$ that breaks the parity symmetry of the potential, and has therefore a strong impact on the Hamiltonian. In order to eliminate the spurious odd-parity term and obtain a $\pi$-periodic element with the negative sign, a possibility is to replace a single kite element by the series of two kite elements \cite{hays2025}, whose weak asymmetries, leading to the presence of both first and second harmonic in the potential [cf. Eq.~\eqref{Eq:kite}], are not detrimental for the qutrit scheme. 

Assuming that an effective $\pi$-periodic element with the correct negative sign is realized, the Hamiltonian describing a qutrit reads 
\begin{equation}
\label{Eq:Hqutrit}
H=-4E_C\partial_\varphi^2+\frac{E_L}{2}(\varphi-\varphi_x)^2+\frac{\pi^2E_L}{4}\cos(\varphi)-E_J\cos(2\varphi),
\end{equation}
where we denoted with $E_J$ the energy of the $\pi$-periodic element. 

The spectrum can be obtained numerically with high accuracy by expanding the Hamiltonian on the basis of eigenstates of the harmonic oscillator constituted by the capacitance and the inductance connected in parallel (see Appendix \ref{App:qutrit-exact-spectrum}). The result is shown in the left panel of Fig.~\ref{fig_qutrit_circuit}(b), where a three-level subspace appears around $\varphi_x=0$, well separated from higher-energy states. Few eigenfunctions are shown in Fig.~\ref{fig:qutrit_main}(b) and  in the right panel of Fig.~\ref{fig_qutrit_circuit}(b), properly displaced at their energy on the potential landscape. Clear fluxon states appear strongly localized at the minima $\varphi_l=\pi l$ of the $-E_J\cos(2\varphi)$ term and weakly hybridized among each other. The parity of the eigenstates is also manifest. The third and fourth excited states are weakly hybridized fluxon states at $l=\pm 2$. The latter are shifted up in energy with respect to the $l=0,\pm 1$ fluxon states by approximately $E_L(l\pi)^2/2$, hybridizing both with one another and with the first excited states (plasmons) in the two wells to the side of the central one. 

We now compare the result of the numerical diagonalization with the effective model  described in Sec.~\ref{Sec:Fluxon-basis} 
and detailed in Appendix~\ref{App:tight_binding}. Considering only the three states localized around $\varphi_l=\pi l$, with $l=0,\pm 1$, the effective Hamiltonian Eq.(\ref{Eq:Hfluxon}) reads
\begin{equation}\label{Eq:qutrit_3x3}
    h=\left(\begin{array}{ccc}
    \varepsilon_{-1} & -t & 0\\
    -t & \varepsilon_0 & -t\\
    0 & -t & \varepsilon_1
    \end{array}
    \right),
\end{equation}
where we have $\varepsilon_l$ and $t$ given by Eq.~\eqref{Eq:varepsilon} and Eq.~\eqref{Eq:t}, respectively, in Appendix \ref{App:tight_binding}. In Fig.~\ref{fig-spect-compare} we compare the numerical spectrum with the one resulting from the low-energy approximation. The main features are well captured and the approximation is valid in the regime $E_L<E_C\ll E_J$, although the tunneling matrix element $t$ appears to be slightly overestimated.

\begin{figure}[t]
	\centering
    \includegraphics[width=1\linewidth]{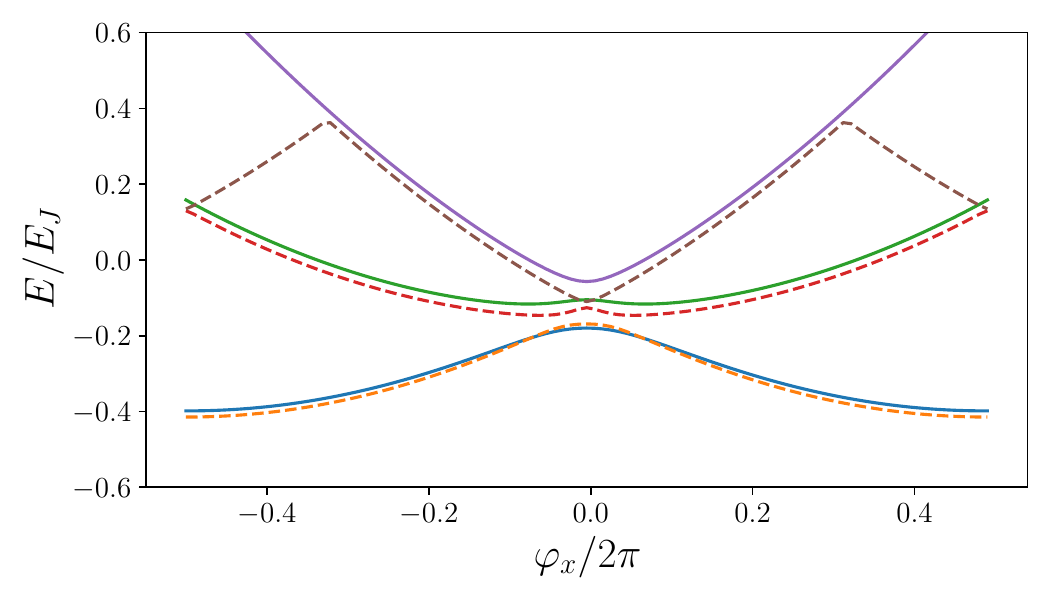}
	\caption{Comparison between the numerically computed qutrit spectrum (dashed line) and the low-energy one provided by Eq.~\eqref{Eq:qutrit_3x3}, as a function of the flux $\varphi_x$ through the main loop. The parameters are $E_L/E_J=0.06$ and $E_C/E_J=0.08$.}
	\label{fig-spect-compare}
\end{figure}

\section{Qutrit selection rules and manipulation}
\label{Sec:qutrit-manipulation}

The effective low-energy description developed in Secs.~\ref{Sec:Fluxon-basis} and \ref{Sec:qutrit} enables us to study the coupling of the system to external microwave radiation and to devise schemes for qutrit manipulation. The fluxonium character of the device suggests the possibility to drive the qutrit transitions via charge and phase dipole interactions. At $\varphi_x=0$, since the potential is even under $\varphi \to -\varphi$, the allowed dipole transition matrix elements can be inferred from the parity of the eigenstates, which alternates starting from the even-parity ground state. Additional constraints on allowed transitions arise from the confinement of the qutrit eigenstates as functions of $\varphi$. In particular, as shown in Fig.~\ref{fig_qutrit_circuit}(b-c), the lowest qutrit states are confined to $-\pi \lesssim  \varphi \lesssim  \pi$ and have negligible overlap with the two higher fluxon states localized around $\varphi = \pm 2\pi$; these two multiplets are therefore only weakly coupled by perturbations that are local in the phase, such as those coupled through the flux. We note that charge and phase operators generically  couple the lowest energy qutrit eigenstates to higher energy plasmon states; the latter are indeed  more delocalized (as function of $\varphi$). 

\subsection{Dipole matrix elements}

\begin{figure}[t]
	\centering
    \includegraphics[width=1\linewidth]{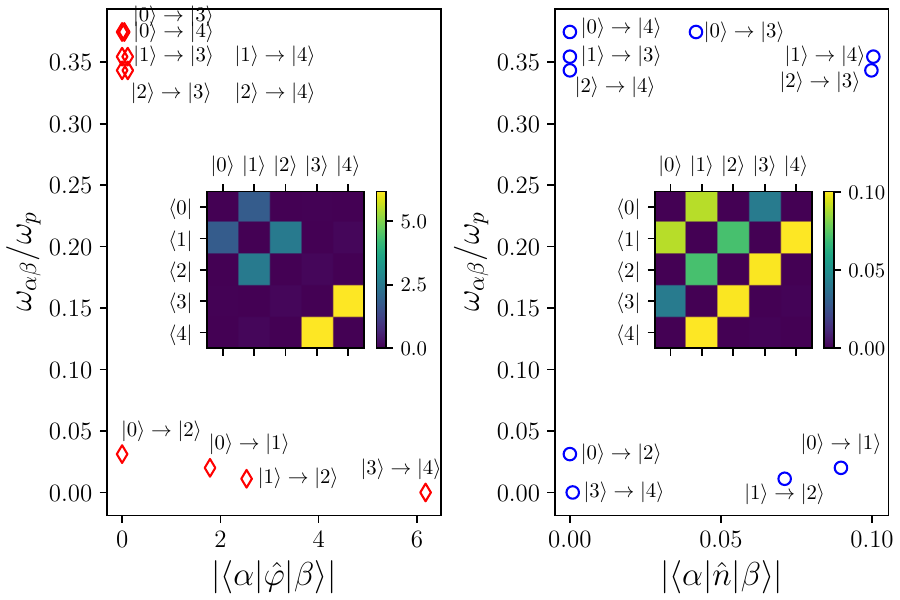}
	\caption{Chart of the dipole matrix elements at zero flux $\varphi_x=0$ of the flux operator $\hat{\varphi}$ (left panel) and of the charge operator $\hat{n}$ (right panel) between the qutrit eigenstates $|\alpha\rangle$ and $|\beta\rangle$, for the first five levels, $\alpha,\beta=0,1,\ldots,4$. Along y axes,  we display the corresponding energy difference $\omega_{\alpha\beta}=|E_\alpha-E_\beta|$ of the transition in units of the plasma frequency $\omega_{p}=2\sqrt{8E_CE_J}$.  The insets are matrix plots of the absolute value of the matrix elements. The calculation assumed the system parameters  $E_C/E_J=0.08$, $E_L/E_J=0.03$.}
	\label{Fig-DME}
\end{figure}

In Fig.~\ref{Fig-DME}, we show the matrix elements of the flux and the charge operators between the eigenstates and the corresponding energy level splitting. Denoting by $|\alpha\rangle$, with $\alpha = 0, 1, 2, \ldots$, the eigenstates with energies $E_\alpha$ of the Hamiltonian in Eq.~\eqref{Eq:Hqutrit}, in Fig.~\ref{Fig-DME} (a) and (b) we plot respectively the phase and charge dipole matrix elements, $\langle\alpha|\hat{\varphi}|\beta\rangle$ and $\langle\alpha|\hat{n}|\beta\rangle$,  and the energy of the corresponding transitions $E_\alpha-E_\beta$. We notice that at $\varphi_x=0$ the phase operator mediates only transitions within the qutrit subspace, while the charge operator yields also transitions from the lower fluxon states, with $\alpha = 0, 1, 2$, to the higher ones having $\alpha = 3,4$.

The Hamiltonian describing the first five levels and their coupling {\sl via} dipole matrix elements thus reads
\begin{eqnarray}
    H & \simeq & \sum_\alpha E_\alpha |\alpha\rangle\langle\alpha|
        + \bigl(\lambda^{\mathrm{f}}_{01}|0\rangle + \lambda^{\mathrm{f}}_{21}|2\rangle +\lambda^{\mathrm{c}}_{41}|4\rangle \bigr)\langle 1|
        + \mathrm{H.c.} \nonumber \\
    &+& \bigl(\lambda^{\mathrm{c}}_{03}|0\rangle + \lambda^{\mathrm{c}}_{23}|2\rangle + \lambda^{\mathrm{f}}_{43}|4\rangle\bigr)\langle 3|
        + \mathrm{H.c.},
\end{eqnarray}
where $\lambda^{\mathrm{f}}_{\alpha\beta}$ and $\lambda^{\mathrm{c}}_{\alpha\beta}$ are the flux and charge dipole matrix elements,
\begin{equation}
    \lambda^{\mathrm{f}}_{\alpha\beta} = \varphi_x E_L \langle\alpha|\hat{\varphi}|\beta\rangle,
    \qquad
    \lambda^{\mathrm{c}}_{\alpha\beta} = 8 E_C n_g \langle\alpha|\hat{n}|\beta\rangle,
\end{equation}
shown in Fig.~\ref{Fig-DME}.

\begin{figure}[t]
	\centering
    \includegraphics[width=1\linewidth]{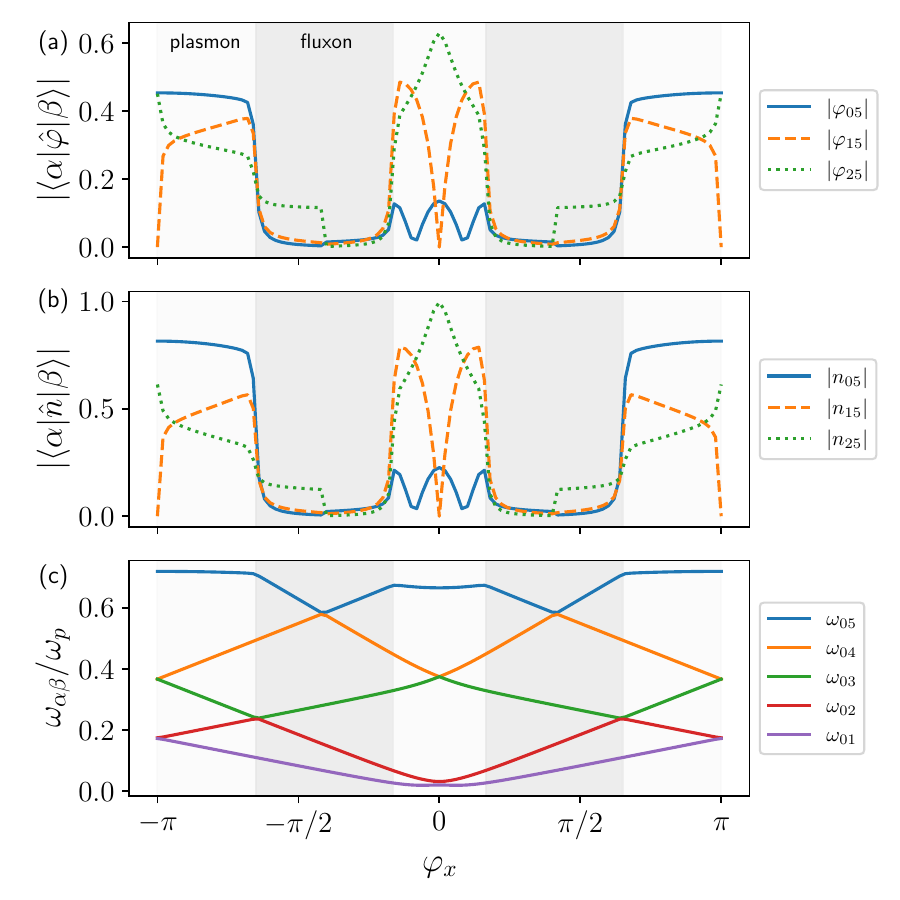}
	\caption{(a) Phase ($\varphi_{\alpha\beta}\equiv \langle\alpha|\hat{\varphi}|\beta\rangle$) and (b) charge ($n_{\alpha\beta}\equiv \langle\alpha|\hat{n}|\beta\rangle$) dipole matrix elements  between the qutrit states $|0\rangle, |1\rangle, |2\rangle$ and the fifth excited state $|5\rangle$ as a function of the flux $\varphi_x$. (c) Energy differences $\omega_{\alpha\beta}\equiv|E_\alpha-E_\beta|$ between the ground state and the first five excited states, in units of $\omega_{p}=2\sqrt{8E_JE_C}$ and as a function of the external flux. The calculation assumed the system parameters  $E_C/E_J=0.08$, $E_L/E_J=0.03$.
    }
	\label{Fig-dipole-flux}
\end{figure}

An analogous analysis can be done for the transitions between the qutrit eigenstates and the higher-energy plasmon states. To simplify the treatment, we only consider the first plasmon state,  $|5\rangle$.  We extend the analysis to finite flux $\varphi_x$, since at $\varphi_x=0$ only the states $|0\rangle$ and $|2\rangle$  couple to $|5\rangle$ via dipole matrix elements.  The corresponding phase and charge dipole-matrix elements are shown in Fig.~\ref{Fig-dipole-flux} (a) and (b), respectively, while  Fig.~\ref{Fig-dipole-flux} (c) shows the transition energies between the first five excited states and the ground state as a function of $\varphi_x$. We notice that the state $|5\rangle$ has a plasmon character only in a range of flux around $\varphi_x=0$ and $\varphi_x=\pm\pi$, where sizable dipole matrix elements arise, accompanied by a weak dependence of the transition on $\varphi_x$. In the other regions, the state $|5\rangle$ has a fluxon-like character.

\subsection{Manipulation}

Depending on the energy scales of the system's parameters, the qutrit level splittings can be either on the order of GHz or on the order of tens of MHz. In the first case, coherent dynamics within the qutrit subspace can be achieved  as in the fluxonium case. In particular, working at the sweet spot $\varphi_x = 0$, transitions $|0\rangle \to |1\rangle$ and $|1\rangle \to |2\rangle$ can be induced by driving the system through the  external flux at frequencies $\omega_{01} = E_1 - E_0$ and $\omega_{12} = E_2 - E_1$. 

\begin{figure}[t]
	\centering
    \includegraphics[width=1\linewidth]{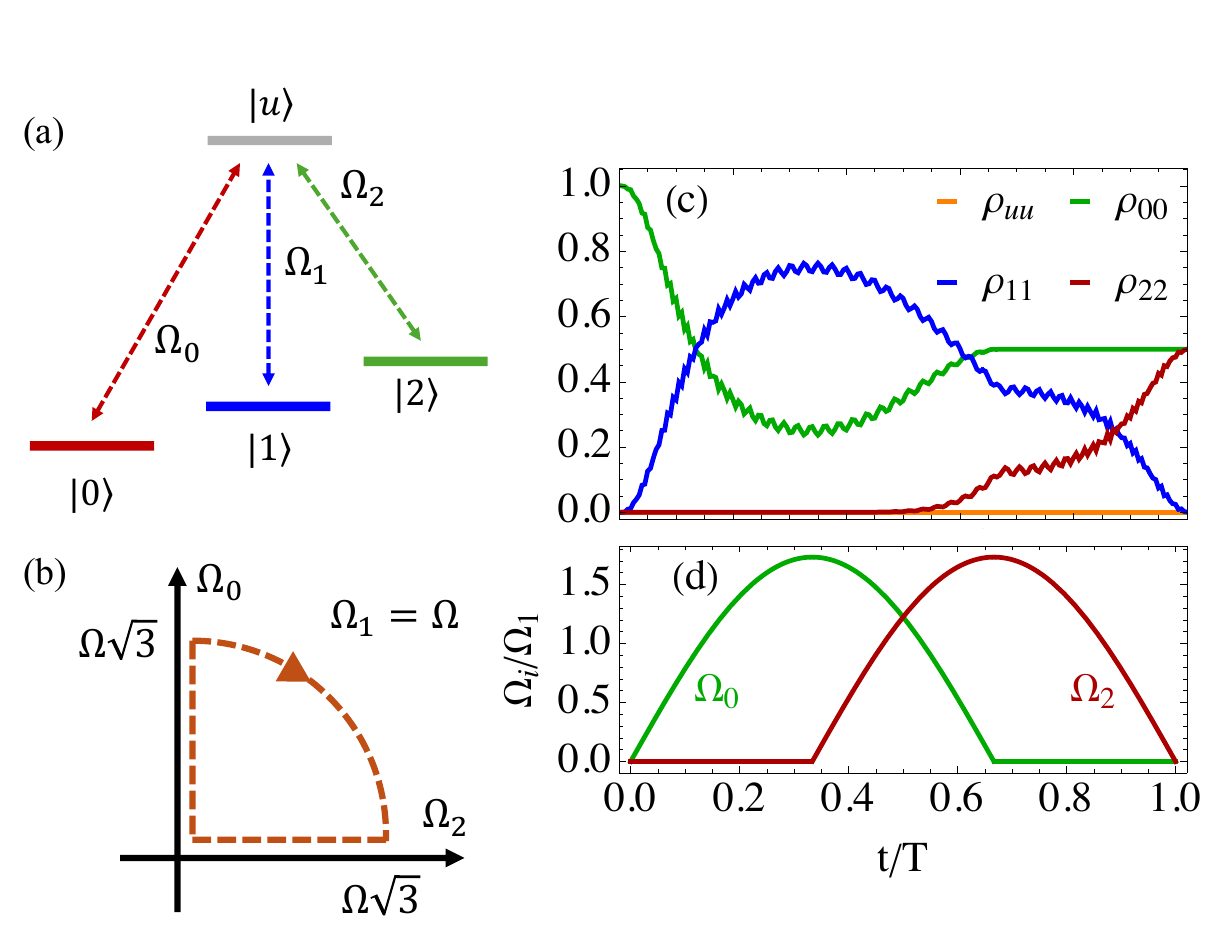}
	\caption{Possible manipulation scheme inspired by the non-Abelian STIRAP protocols. (a) Structure of the Rabi Hamiltonian. (b) and (d) Possible adiabatic cycle implementing a $\pi/2$ rotation between the states $|0\rangle$ and $|2\rangle$. (c) Numerical evolution of the populations of the different levels within the rotating wave approximation for $T=500/\Omega_1$; $\Omega_1$ is kept constant during the whole cycle.}
	\label{Fig-stirap}
\end{figure}

When the qutrit frequencies fall in the MHz range, transitions within the qutrit subspace can instead be induced by exploiting a non-computational upper level, which we denote as $|u\rangle$. There are mainly two possibilities for the state $|u\rangle$: it can be one of the upper fluxon-like state, such as $|3\rangle$, or it can coincide with the lowest plasmon level $|5\rangle$. In both cases,  at the sweet spot $\varphi_x=0$ the dipole transition $|1\rangle\to|u\rangle$ is forbidden by symmetry, since  the states $|1\rangle$, $|3\rangle$ and $|5\rangle$ are all odd. Furthermore, the quasi-degeneracy of the qutrit levels  may prevent selectively addressing the transitions.

To overcome these limitations, several solutions can be employed: analogously to what is done for the fluxonium, we can break the parity-enforced selection rules  through a third-order nonlinear element as proposed in Ref.~\cite{vool2018}; or we can exploit two-photon processes~\cite{nesterov2021}. Alternatively, in this work, we choose to detune the qutrit away from the flux sweet spot~\cite{manucharyan2009}, so to distort the symmetry of level $|5\rangle$ and induce a hybridization between $|3\rangle$ and $|4\rangle$. The driving protocols traditionally employed for fluxonium qubits \cite{manucharyan2009,novikov2016,earnest2018,vool2018,nesterov2021} can be then extended to the qutrit. These include resonant sequential driving and $\Lambda$-type driving schemes. By doing so, the qutrit levels $|0\rangle$, $|1\rangle$, $|2\rangle$ are each coupled to $|u\rangle$, realizing a tripod level structure. Thanks to its extended character as a function of the phase, level $|5\rangle$ grants a more robust coupling, we will thus focus on this level identifying $|u\rangle$ with  $|5\rangle$.

\subsubsection{Adiabatic tripod scheme}

Within the Rotating Wave Approximation (RWA), the low-energy Hamiltonian in presence of driving has the following structure:
\begin{equation}
H_{\rm RWA}=\sum_{\alpha=0,1,2}\left(\Omega_\alpha(t) |u\rangle\langle\alpha|+{\rm H.c.}\right),
\end{equation}
where $\Omega_\alpha(t)$ indicate the Rabi amplitudes corresponding to the transitions $|\alpha\rangle \rightarrow |u\rangle$. The resulting tripod level structure paves the way to stimulated Raman adiabatic passage (STIRAP) approaches~\cite{vitanov2017}, including the non-Abelian STIRAP proposed {\sl e.g.} in Refs.~\cite{unanyan1999,faoro2003,brosco2021}. As an illustration, Fig.~\ref{Fig-stirap} shows an example of a STIRAP protocol on the qutrit implementing a $\pi/2$ rotation from the ground state $|0\rangle$ to the superposition $\frac{1}{\sqrt{2}}(|0\rangle+|2\rangle)$. The protocol relies on the adiabatic modulation of two resonant driving amplitudes while keeping the amplitude of drive $|1\rangle\rightarrow |u\rangle$ constant, and it is geometrically protected, robust against small variations of the driving parameters and dynamical perturbations \cite{brosco2021}. We notice that the peculiar level structure of the qutrit enables to fulfill simultaneously the adiabaticity condition and the RWA. The former requires the time scale of the modulation of the Rabi driving $T$ to be much larger than the energy gap in the rotating frame $\bar \Omega=\sqrt{\Omega_1^2+\Omega_2^2+\Omega_3^2}$ {\sl i.e.} $\bar \Omega\, T\gg 1$; while the RWA requires the Rabi amplitude to be much smaller than the driving frequencies, {\sl i.e.} $\Omega_i\ll |E_u-E_i|$. Both these assumptions can be fulfilled for realistic values of the qutrit parameters. The STIRAP scheme illustrated above can be utilized to initialize and to a certain extent manipulate the qutrit state; its main advantage is the geometric nature, which makes it robust and at the same time easy to engineer, as change in the geometric structure allows to change the rotation angle. On the other hand, adiabaticity implies a low gate speed; this drawback could be overcome by optimizing the driving pulse, as demonstrated in a number of works, see {e.g.} Refs.~\cite{florio2006,sjoqvist2012,giannelli2022,zheng2022}. The adiabatic protocol described can be extended to the general case of a qudit \cite{vitanov2017} and to the manipulation of coupled qudits \cite{faoro2003,danieli2026}.

\section{Conclusions}

In this work, we have introduced a novel superconducting circuit architecture for qudit encoding, conceptually extending the fluxonium design to $d$-dimensional Hilbert spaces. Motivated by the need for more efficient quantum error correction and the inherent limitations of transmon-based qutrits -- specifically their susceptibility to leakage errors -- our proposal leverages a potential energy landscape with $d$ degenerate minima. This structure ensures a large spectral gap that isolates the computational subspace, providing a natural protection mechanism that is missing in current state-of-the-art implementations.

A central feature of our design is the promotion of fractional fluxon states, or {\it fraxons}. By engineered compensation of the inductive energy, we generate a potential where the localized flux corresponds to fractional units of the superconducting flux quantum. We demonstrated that in the regime of high flux localization, the system is accurately described by a tight-binding model, where quantum phase-slip processes drive the hopping between minima and determine the level splitting.

Although the  high-harmonic terms required in the Josephson potential are not standard in conventional junctions, we present a scalable ``recipe'' to generate them through an iterative scheme. This procedure can be applied to different platforms: while we focused on Josephson tunnel junctions, the general framework is readily applicable to emerging platforms such as semiconductor-superconductor hybrids~\cite{feldstein-bofill2026}, twisted d-wave superconductors~\cite{confalone2025}, or any system offering high-degree control over harmonic content.

Through a detailed analysis of the $d=3,4, 5$ cases, we highlighted the versatility of the physical scheme for quantum gate implementation. For the specific case of a qutrit, we demonstrated that the well-isolated low-energy subspace is ideally suited for adiabatic manipulation. By employing a tripod coupling structure, we showed that single-qutrit gates can be performed via non-Abelian stimulated Raman adiabatic passage (STIRAP). 

The tripod structure offers a unique opportunity to implement holonomic quantum computing, where information is protected by the geometric nature of the gates \cite{wilczek1984,zanardi1999,duan2001}. In addition, $d$-level systems have been considered also for the realization of spin coherent \cite{dooley2013,joo2016,nguyen2024,denis2026} and GHZ-like states~\cite{shen2021}. Furthermore, the ability to engineer high-spin systems and simulate high-energy phenomena in a superconducting circuit marks a significant step forward in quantum engineering \cite{senko2015,wang2023,blok2021,gustafson2022,spagnoli2025,wang2020}. Finally, the parallels between our circuit and atomtronic systems -- specifically cold atoms in ring-shaped potentials -- suggest that the principles of fractional fluxon engineering could find fruitful application in neutral superfluid circuits \cite{amico2014superfluid,aghamalyan2016atomtronic,amico2022colloquium}.

\acknowledgments

L.C. acknowledges the Fondazione Cariplo under the grant 2023-
2594. U.V. was partially supported by the European Union (ERC-StG, cQEDscope, 101075962), V.B. acknowledges support from  Project PNRR MUR PE 0000023-NQSTI  and by the Deutsche Forschungsgemeinschaft (DFG 539383397).

\appendix

\section{Fourier potential engineering}
\label{App:potential_engineering}

Here, we provide details for generating the potential that gives rise to the generalized fluxonium featuring $d$ degenerate minima.  For the case of odd-$d$, the natural ansatz is a potential of the form
\begin{equation}
    U(\varphi)=\frac{E_L}{2}\varphi^2-E_J\cos(L_d\varphi)+E_L\sum_{n<L_d}a_n\cos(n\varphi).
\end{equation}
Clearly, being $L_d$ the largest frequency in the harmonic expansion, the sum over $n$ is restricted to harmonics such that $n<L_d$. A requirement to have a well defined system of equations for the coefficients $a_n$ is that the basis calculated at approximate minima $\varphi_l=2\pi l/L_d$ is composed by all different terms, that is to have all different $\cos(2\pi n/L_d)$. This is accomplished by choosing $0<n/L_d<1/2$. 

To proceed further, we assume that $\eta\equiv E_L/E_J\ll 1$ and postulate a series expansion in $\eta$ for the location of the minima $\varphi_l$ and the value of the coefficients $a_n$, such that
\begin{eqnarray}
    \varphi_l&=&\frac{2\pi l}{L_d}+\eta \delta\varphi_l^{(1)}+\eta^2\delta\varphi_l^{(2)}+\ldots\\
    a_n&=&a_n^{(0)}+\eta a_n^{(1)}+\eta^2a_n^{(2)}+\ldots.
\end{eqnarray}
To have $d$-degenerate minima we then need to simultaneously solve $U(0)=U(\varphi_l)$ and $U'(\varphi_l)=0$.  This way, we obtain a linear system of equations for $\delta\varphi_l^{(j)}$ and $a_n^{(j)}$. By setting to zero the coefficients of the expansion of $U(0)=U(\varphi_l)$ proportional to $\eta$ (the one at zeroth order is automatically satisfied), we find that the $a^{(0)}_n$  must satisfy the linear system
\begin{equation}
    \sum_{n=1}^{L_d/2}\sin^2\left(\frac{\pi}{L_d}nl\right)a^{(0)}_n=\left(\frac{2\pi}{L_d}\right)^2\frac{l^2}{4},
\end{equation}
that is solved by inverting the matrix whose components are $M_{nl}=\sin^2\left(\frac{\pi}{L_d}nl\right)$, for $n,l=1,\ldots,L_d/2$. Analogously, by setting to zero the coefficient proportional to $\eta$ of the expansion of $U'(\varphi_l)=0$ we obtain
\begin{equation}\label{eq:dphil1}
\delta\varphi_l^{(1)}=-\frac{1}{L^2_d}\left[\frac{2\pi l}{L_d}-\sum_{n=1}^{L_d/2}na^{(0)}_n\sin\left(\frac{2\pi l n}{L_d}\right)\right].
\end{equation}
Proceeding analogously, by setting to zero the coefficient proportional to $\eta^2$ of the expansion of $U(0)=U(\varphi_l)$ we find that the $a^{(1)}_n$ must satisfy the linear system  
\begin{equation}
    \sum_{n=1}^{L_d/2} a^{(1)}_nM_{nl} = -\frac{1}{4}L_d^2\left(\delta\varphi^{(1)}_l\right)^2.
\end{equation}
For simplicity, we only consider coefficients $a_n$ at zeroth order in $\eta$ [except in Eq.~\eqref{Eq:qutrit-potential-general}]. A similar procedure can be employed for the case of even $d$.

\section{Energy-phase relation of the modular element}
\label{App:energy-phase-expansion}

Here, we provide details of the energy-phase relation describing the modular element. We start by the potential describing a series of an inductance with energy $E_L$ and a Josephson junction with energy $E_J$, 
\begin{equation}
    E_J(\varphi,\delta\varphi)=-E_J\cos[(\varphi+\delta\varphi)/2]+E_L(\varphi-\delta\varphi)^2/8
\end{equation}
Minimization of the potential for $\delta\varphi$ yields the equation
\begin{equation}
    \delta\varphi=\varphi-\frac{2E_J}{E_L}\sin[(\varphi+\delta\varphi)/2].
\end{equation}
For $\tau\equiv E_J/E_L\ll 1$, we can attempt an expansion of $\delta\varphi$ in powers of $\tau$. We express $\delta\varphi=\varphi-2\tau z$, so that we need to solve $z=\sin(\varphi-\tau z)$. Assuming $z=z_0+\tau z_1+\tau^2 z_2+\ldots$,  we find
\begin{equation}
\delta\varphi=\varphi-2\tau\sin(\varphi)+\tau^2\sin(2\varphi)+\ldots
\end{equation}
and by plugging the result in the potential we find
\begin{eqnarray}
    \frac{E_J(\varphi)}{E_L}&=&-\tau\left(1-\frac{\tau^2}{8}\right)\cos(\varphi)\nonumber\\
    &+&\frac{\tau^2}{4}\cos(2\varphi)-\frac{\tau^3}{8}\cos(3\varphi)+\ldots.,
\end{eqnarray}
The alternation in sign is analogous to the expansion of an Andreev state in terms of the transparency $\tau$ of the conduction channels  \cite{willsch2024,chirolli2025}.

\section{Qudit effective low-energy model}
\label{App:tight_binding}

Here, we give more details on the derivation of the low-energy model outlined in Sec.~\ref{Sec:Fluxon-basis}. 
First of all, after solving for the coefficients $a_n$ of the potential at $\varphi_x=0$ as explained in Appendix~\ref{App:potential_engineering}, we expand the potential for $\varphi_x\neq0$ around the minima where fractional fluxon states are confined. We focus on the odd-$d$ case for which $L_d=d-1$ and follow the procedure outlined in Ref.~\cite{catelani2011}. The full potential energy 
\begin{eqnarray}
    U(\varphi,\varphi_x)&=&\frac{E_L}{2}(\varphi-\varphi_x)^2-E_J\cos(L_d\varphi)\nonumber\\
    &+&E_L\sum_{n=1}^{L_d/2}a_n\cos(n\varphi),
\end{eqnarray}
is extremized by values of the phase $\varphi_l$ satisfying
\begin{equation}
    E_L(\varphi_l-\varphi_x)+L_dE_J\sin(L_d\varphi_l)-E_L\sum_{n=1}^{L_d/2}n a_n\sin(n\varphi_l)=0.
\end{equation}
For $\eta\equiv E_L/E_J\ll 1$, we can solve for $\varphi_l$ at leading order in $\eta$ and find [cf. Eq.~\eqref{eq:dphil1}]
\begin{equation}
    \varphi_l=\frac{2\pi l}{L_d}
    -\frac{\eta}{L_d^2} \left(\frac{2\pi l}{L_d}-\varphi_x\right)+\frac{\eta}{L_d^2}\sum_n n a_n \sin\left(\frac{2\pi nl}{L_d}\right).
\end{equation}
Around each minimum $\varphi_l$ we expand the potential to second order as 
\begin{equation}\label{Eq:expMinima}
    H_l=-4E_C\partial^2_\varphi+\frac{\bar{E}_{L,d}}{2}(\varphi-\varphi_l)^2+U(\varphi_l,\varphi_x),
\end{equation}
with the effective inductance 
\begin{equation}
    \bar{E}_{L,d}=E_L+L_d^2E_J\cos(L_d\varphi_l)-E_L\sum_{n=1}^{L_d/2} a_nn^2\cos(n\varphi_l),
\end{equation}
that has a weak dependence on $l$, and obtain approximate eigenstates
\begin{equation}\label{eq:psil}
    \psi_l(\varphi)=\frac{1}{\sqrt{\ell_d\sqrt{\pi}}}e^{-(\varphi-\varphi_l)^2/2\ell_d^2},
\end{equation}
with energy
\begin{eqnarray}\label{Eq:varepsilon0}
    \varepsilon_l&=&U(\varphi_l,\varphi_x)+\frac{1}{2}\sqrt{8E_C\bar{E}_{L,d}},\\
    &=&\frac{E_L}{2}(\varphi_l-\varphi_x)^2-\frac{E_L}{2}\varphi_l^2+E_0,
\end{eqnarray}
where $\ell_d=(8E_C/\bar{E}_{L,d})^{1/4}$ and $E_0=\frac{1}{2}\sqrt{8E_C\bar{E}_{L,d}}+U(\varphi_l,0)$, that has a weak dependence on $l$.

\begin{figure}[t]
	\centering
    \includegraphics[width=1\linewidth]{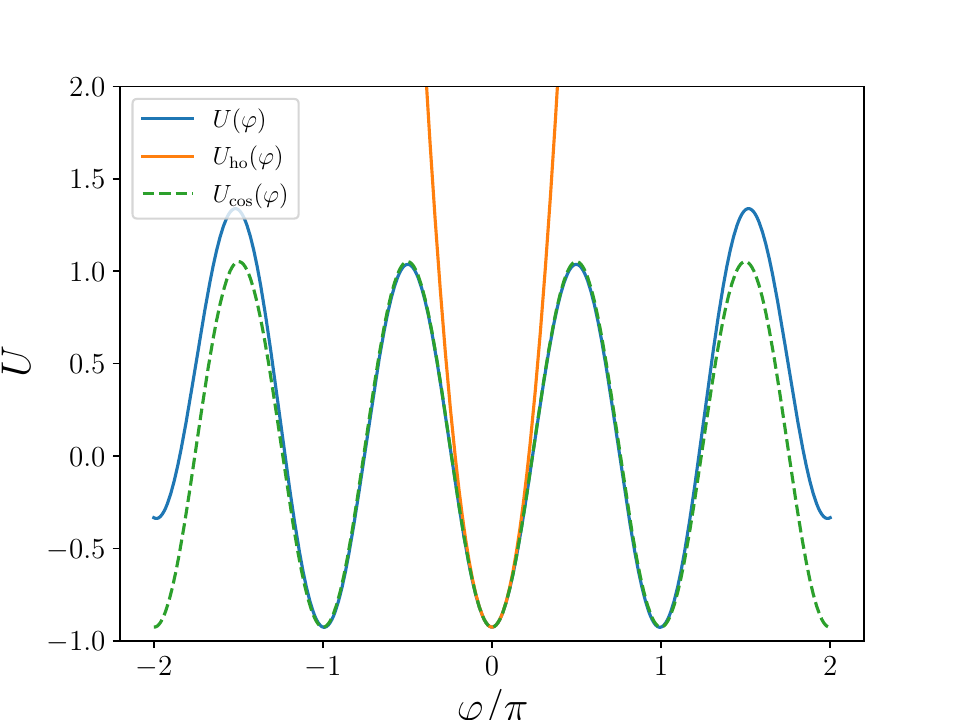}
	\caption{Comparison of the potential realizing a qutrit, its quadratic expansion around $\varphi=0$, and the effective cosine potential $\bar{E}_J[1-\cos(L\varphi)]$.}
	\label{fig-pot-comparisoc}
\end{figure}

The states $\psi_l$ are well-localized around the minima $\varphi_l$ as long as $\ell_d\ll 1$ and we adopt a tight-binding method to estimate the splitting due to tunneling through the barriers. Despite the fact that the barriers between the minima $\varphi_l$ have not equal height, as clear by inspection of Fig.~\ref{fig:potential-qudit}(a), we approximate the potential in the region of the first $d$ minima with $-\bar{E}_J\cos(L\varphi)$, where we choose an effective inverse period $L$, such that $\left.L\varphi_{l}\right|_{l=1,\varphi_x=0}=2\pi$, that is
\begin{equation}
    L=L_d\left[1
    -\frac{\eta}{L_d^2} +\frac{\eta}{2\pi L_d}\sum_na_n n\sin\left(\frac{2\pi n}{L_d}\right)\right]^{-1}.
\end{equation}
and we set
\begin{equation}
    \bar{E}_J=\frac{\bar{E}_{L,d}}{L^2},
\end{equation}
such that the approximate Hamiltonian describing the lowest $d$ minima reads
\begin{equation}
    H_{\rm eff}=-4E_C\partial^2_\varphi+\bar{E}_J\left[1-\cos(L\varphi)\right]+U_0,
\end{equation}
whose expansion around the minima reproduces Eq.~\eqref{Eq:expMinima}, with $U_0=U(\varphi_l,0)$ for $l=0$ and $\varphi_x=0$. The effective cosine potential is shown in Fig.~\ref{fig-pot-comparisoc}, together with the qutrit potential and its quadratic expansion around $\varphi=0$. 
Finally, upon defining $\theta=L\varphi$, we map the problem to the transmon Hamiltonian
\begin{equation}
    H_{\rm Transmon}=-4\bar{E}_C\partial^2_\theta+\bar{E}_J\left[1-\cos(\theta)\right]+U_0,
\end{equation}
with $\bar{E}_C=L^2E_C$. 

\subsection{Tight-binding model}
\label{App:tight-binding}

Following \cite{catelani2011}, we can model the dynamics between the fluxon states localized in the minima of the effective transmon potential through a tight-binding approach. The fluxon states are non-orthogonal, and a correct diagonalization needs to take the overlap matrix into account, such that eigenvalues and eigenstates can be obtained by diagonalizing the matrix $S^{-1/2}H_{\rm qudit}S^{-1/2}$, with $S$ being the matrix of overlap elements $S_{ll'}=\langle l|l'\rangle$. Nevertheless, the overlaps between different fluxon states are strongly suppressed, so that we approximate $S_{l,l'}=\delta_{l,l'}+s\delta_{l,l'+1}+s\delta_{l+1,l'}$ and neglect corrections of order $s\ll1$. Retaining only nearest neighbor tunneling, the tunneling matrix elements between adjacent minima can be written as \cite{catelani2011,connor1984}
\begin{equation}
    t=2\sqrt{\frac{2}{\pi}}\hbar\omega_p\left(\frac{8\bar{E}_J}{\bar{E}_C}\right)^{1/4}e^{-\sqrt{8\bar{E}_J/\bar{E}_C}}.
\end{equation} 
Notice that, as previously anticipated, we neglect a weak modulation of the potential that cannot be exactly canceled, such that we approximate $t_{l,l+1}=t$. We can then construct an effective Hamiltonian, that for the $d=5$ case reads
\begin{equation}
    h=\left(\begin{array}{ccccc}
    \varepsilon_{-2} & -t & 0 & 0 & 0\\
    -t & \varepsilon_{-1} & -t & 0 & 0\\
    0 & -t & \varepsilon_0 & -t & 0\\
    0 & 0 & -t & \varepsilon_1 & -t\\
    0 & 0 & 0 & -t & \varepsilon_2
    \end{array}
    \right).
\end{equation}
As remarked above, the $t$'s are in principle not all equal, since the potential barriers are slightly different from each other, but we ignore this difference. We can obtain a better approximation than Eq.~\eqref{Eq:varepsilon0} for $\varepsilon_l$ by taking the expectation value of the full potential on the localized states [Eq.~\eqref{eq:psil}],
\begin{align}
    \varepsilon_l&= \frac{1}{2}\sqrt{8E_C\bar{E}_{L,d}}+\langle \psi_l|U(\varphi, \varphi_x)-\frac{\bar{E}_{L,d}}{2}(\varphi-\varphi_l)^2|\psi_l\rangle\\
    &= \frac{1}{2}\sqrt{8E_C\bar{E}_{L,d}}+\frac{E_L}{4}\left[\ell_d^2 +2(\varphi_l-\varphi_x)^2\right]-\frac{\ell_d^2}{4}\bar{E}_{L,d}\nonumber\\
    &- E_J e^{-L_d^2\ell_d^2/4}\cos(L_d\varphi_l)+\sum_n a_ne^{-n^2\ell_d^2/4}\cos(n\varphi_l).
\end{align}

For the qutrit case we have \begin{eqnarray}
\varphi_l&=&\pi l-(E_L/E_J)\left(\pi l-\varphi_x\right)/4, \label{eq:pl3}\\
\bar{E}_{L}&=&E_L+4E_J\cos(2\varphi_l)-E_L\frac{\pi^2}{4
}\cos(\varphi_l),\\
\varepsilon_l&=&\frac{1}{2}\sqrt{8E_C\bar{E}_{L}}+\frac{E_L}{4}(\ell^2 +2(\varphi_l-\varphi_x)^2)-\frac{\ell^2}{4}\bar{E}_L\nonumber\\
    &-&E_J e^{-\ell^2}\cos(2\varphi_l)+\frac{\pi^2E_L}{4}e^{-\ell^2/4}\cos(\varphi_l).\label{Eq:varepsilon}\\
t&=&2\sqrt{\frac{2}{\pi}}\sqrt{8E_C\bar{E}_{L}}\left(\frac{8\bar{E}_L}{L^4E_C}\right)^{1/4}e^{-\sqrt{8\bar{E}_L/L^4E_C}}\label{Eq:t}
\end{eqnarray}
with $\ell=(8E_C/\bar{E}_L)^{1/4}$ and $L=2/(1-\eta/4)$.

\section{Numerical spectrum}
\label{App:qutrit-exact-spectrum}

Here, we provide details concerning the numerical calculation of the spectrum of the system and a low-energy effective model describing the qutrit subspace. We define $H_{\rm ho}=-4E_C\partial_\varphi^2+E_L\varphi^2/2$ and introduce a basis of Fock states $|n\rangle$ for which $\langle m|H_{\rm ho}|n\rangle=\delta_{nm}\sqrt{8E_CE_L}(n+1/2)$. The dependence on the external flux $\varphi_x$ can be gauged to the Josephson potential, such that $U_d(\varphi)\to U_d(\varphi+\varphi_x)$. Introducing $\sigma=(8E_C/E_L)^{1/4}$ and
\begin{equation}
    D_{mn}(z)\!
    =\!\begin{cases}
    \sqrt{\frac{n!}{m!}}z^{m-n}e^{-|z|^2/2}L_n^{(m-n)}(|z|^2), & m\geq n,\\
    \sqrt{\frac{m!}{n!}}(-z^*)^{n-m}e^{-|z|^2/2}L_m^{(n-m)}(|z|^2), & m\leq n,
  \end{cases}\nonumber
\end{equation}
that are the matrix elements of the displacement operator $\hat{D}(z)=e^{z a^\dag-z^* a}$, $D_{mn}(z)=\langle m|e^{z a^\dag-z^* a}|n\rangle$, with $L_n^{(k)}$ generalized Laguerre polynomials, we can calculate the matrix elements of the higher harmonics appearing in the general Hamiltonian Eq.~\eqref{Eq:Hqudit} describing qudits,
\begin{equation}
    \langle m|e^{ip_n(\varphi+\varphi_x)}|n\rangle=e^{ip_n\varphi_x+i\frac{\pi}{2}(m-n)}D_{mn}\left(\frac{p_n\sigma}{\sqrt 2}\right).
\end{equation}
Furthermore, the matrix elements of charge operator $\hat{n}=-i\partial_\varphi$ and flux operator in the Fock basis $|n\rangle$ read 
\begin{eqnarray}
    \langle m|\hat\varphi|n\rangle&=&\frac{\sigma}{\sqrt 2}(\sqrt{n}\delta_{n-1,m}+\sqrt{n+1}\delta_{n+1,m}),\\
    \langle m|\hat{n}|n\rangle&=&\frac{-i}{\sigma\sqrt{2}}(\sqrt{n}\delta_{n-1,m}-\sqrt{n+1}\delta_{n+1,m}).
\end{eqnarray}
The matrix elements of the full qutrit Hamiltonian Eq.~\eqref{Eq:Hqutrit} in this basis read
\begin{eqnarray}
    H_{mn}&=&\delta_{nm}\sqrt{8E_CE_L}(n+1/2)\nonumber\\
    &+&\frac{\pi^2 E_L}{4}D_{mn}(\sigma/\sqrt{2})\cos[\varphi_x-\pi(m-n)/2]\nonumber\\
    &-&E_JD_{mn}(\sqrt{2}\sigma)\cos[2\varphi_x-\pi(m-n)/2],
\end{eqnarray}
In the working regime $E_J\gg E_C,E_L$, the level spacing provided by $\sqrt{8E_CE_L}$ is small compared to $E_J$ and a large number of Fock states is needed to describe well the qutrit low-energy spectrum. In Fig.~\ref{Fig-DME} we employ about 100 Fock states.

In Fig.~\ref{Fig-DME} we plot the dipole matrix elements of the flux and the charge operators, calculated on the eigenstates at zero external flux. Whereas the flux operator has sizable matrix elements only between the higher energy states $|3\rangle$ and $|4\rangle$, and a weak contribution between the pairs $\{|0\rangle,|1\rangle\}$ and $\{|1\rangle,|2\rangle\}$, the charge operator has a sizable matrix element for all transitions, except for the $\{|3\rangle,|4\rangle\}$ one.

\bibliography{qutrit-bib}{}

\end{document}